\documentclass[preprint,showpacs,preprintnumbers,amsmath,amssymb,nofootinbib]{revtex4}
\usepackage{epsfig}

\newcommand{\be}{\begin{equation}}
\newcommand{\ee}{\end{equation}}
\newcommand{\beqq}{\setlength\arraycolsep{2pt}\begin{eqnarray}}
\newcommand{\eeqq}{\vspace{0cm} \end{eqnarray}}
\newcommand{\bea}{\begin{eqnarray}}
\newcommand{\eea}{\end{eqnarray}}
\newcommand{\hsp}{\hspace{1cm}}

\begin{document}

\title{Particle creation in a $f(R)$ theory with cosmological constraints}

\author{S. H. Pereira} \email{shpereira@gmail.com}

\affiliation{UNESP - Universidade Estadual Paulista -- Campus de Guaratinguet\'a - DFQ \\ Av. Dr. Ariberto Pereira da Cunha, 333 -- Pedregulho\\
12516-410 -- Guaratinguet\'a, SP, Brazil}

\author{R. F. L. Holanda}\email{holanda@uepb.edu.br}

\affiliation{Universidade Estadual da Para\'iba -- Departamento de F\'isica\\
Rua Bara\'unas, 351 -- Bairro Universit\'ario \\ 58429-500 -- Campina Grande, PB, Brazil}

\affiliation{Universidade Federal de Campina Grande -- Departamento de F\'isica\\
R. Apr\'igio Veloso, 882 - Bodocongo, 351 -- Bairro Universit\'ario \\ 58109-900 -- Campina Grande, PB, Brazil}


\begin{abstract}In this paper we study the creation of super-massive real scalar particles in the framework of a $f(R)=R-\beta/R^n$ modified gravity theory, with parameters constrained by observational data. The analysis is restrict to a homogeneous and isotropic flat and radiation dominated universe. We compare the results to the standard Einstein gravity with cosmological constant ($\Lambda CDM$ model), and we show that the total number density of created particles in the $f(R)$ model is very close to the standard case. Another interesting result is that the spectrum of created particles is $\beta$ independent at early times.
\end{abstract}

\maketitle

\section{Introduction}

 Recently, non-standard gravity theories have been proposed as an alternative to understand the physical mechanism behind the late-time acceleration of the universe. This is naturally obtained in $f(R)$ gravity theories (for a review, see \cite{sotiri}). In such an approach, the curvature scalar  $R$ in the Einstein-Hilbert action is replaced by a general function $f(R)$, so that the standard Einstein field equation is recovered as a particular case \cite{fR,allemandi,allemandi2,sotiri2,meng,dolgov,cembranos,gaspe,vilk}. A review of various modified gravities considered as gravitational alternative for dark energy, an unified description of the early-time inflation and the solutions of some related problems in $f(R)$ theories have been presented by Nojiri and Odintsov \cite{nojiri}. 

Cosmological constraints on $f(R)$ gravity theories has also been studied recently \cite{carvalho,elgaroy,fay,koivisto}. Of particular interest is the model $f(R)=R-\beta/R^n$, investigated by Carvalho et al. 2008 \cite{carvalho}. They used determinations of the Hubble function, H(z), based on differential age method to place bounds on the free parameters $n$ and $\beta$, combined with constraints from Baryon Acoustic Oscillations (BAO) and Cosmic Microwave Background (CMB) measurements. The best-fit values for $n$ and $\beta$ obtained from H(z)+BAO+CMB test are $0.03$ and $4.7$, respectively, with the parameters $n$ to lie in the intervals $n \in [-0.25,0.35]$ and $\beta \in [2.3,7.1]$ at 99.7\% c.l. The standard $\Lambda CDM$ model corresponds to $n=0$ and $\beta = 4.38$.

By considering a radiation dominated universe, a recent work \cite{saulocarlos} studied the quantum process of particle creation in a $f(R)=R+\beta R^n$ theory. Although very similar to the $f(R)$ considered by Carvalho et al. 2008 \cite{carvalho}, the authors analyzed  only  positive $\beta$ values and for some specific values of $n$, namely $n= (2+2j)/(2j+1)$ and $n= 2j/(2j+1)$, $j=1,2,3,\dots $. It was shown that both massive and massless scalar particles can be produced by purely expanding effects.

In the present work we investigate the particle production in the model $f(R) = R -\beta/ R^n$, with the  parameters obtained by Carvalho et al. 2008 \cite{carvalho}, $n \in [-0.25,0.35]$ and $\beta \in [2.3,7.1]$ at 99.7\% c.l., in a homogeneous and isotropic flat and radiation dominated universe. We show that super-massive particles can be created at the early times, when the universe is radiation dominated. This could be an alternative mechanism to create dark matter particles in the course of the evolution of the universe. Moreover, such $f(R)$ naturally provide an accelerated stage for the evolution of the universe, without to appeal to a dark energy component.

\section{General theory of scalar particle creation}

The phenomenon of particle creation in an expanding universe has been studied by several authors \cite{davies,fulling,grib,mukh2,partcreation,staro,pavlov01,pavlov02,gribmama02,fabris01}. One of the most interesting results  is that in a radiation dominated universe there is exactly no creation of massless particles \cite{parker}, either of zero or non-zero spin. In the description of particle production by the gravitational field, a widely used method is that of instantaneous Hamiltonian diagonalization \cite{pavlov01} suggested by Grib and Mamayev \cite{gribmama02}. Of particular interest are the works relating the gravitational particle production at the end of inflation as a possible mechanism to produce super-heavy particles of dark matter \cite{pavlov02,superDM}.

The canonical quantization of a real minimally coupled scalar field in  curved backgrounds follows in straight analogy with the quantization in a flat Minkowski background. The gravitational metric is treated as a classical external field which is generally non-homogeneous and non-stationary. The basic equation to study  scalar particle creation in a spatially flat Friedmann-Robertson-Walker geometry is \cite{mukh2}:
\begin{equation}\label{nv4}
\chi''_k(\eta)+\omega_k^2(\eta)\chi_k(\eta)=0\,,
\end{equation}
with
\begin{equation}\label{m68}
\omega_k^2(\eta)\equiv k^2+M^2_{eff} \hspace{1cm} \textrm{and} \hspace{1cm} M^2_{eff}\equiv M^2a^2(\eta)-{a''(\eta)\over a(\eta)}\,,
\end{equation}
where $k$ is the Fourier mode or wavenumber of the particle, $\omega_k$ and $M_{eff}$ represents the frequency and the effective mass of the particle, respectively. $a(\eta)$ is the cosmological scale factor in terms of the conformal time $\eta$ and the prime denotes derivatives with respect to it. The conformal time $\eta$ and the physical time $t$ are related by
\be
\eta \equiv \int {dt\over a(t)}\,.\label{conf}
\ee
 
Following standard lines, the quantization can be carried out by imposing equal-time commutation relations to the scalar field $\chi$ and its canonically conjugate momentum $\pi\equiv\chi'$, namely $[\chi(x,\eta) \,, \pi(y,\eta)]=i\delta(x-y)$, and by implementing secondary quantization. After convenient Bogoliubov transformations, one obtains the transition amplitudes to the vacuum state and the associated spectrum of the produced particles in a non-stationary background \cite{grib,mukh2}.

 Usually, the calculations of particle production  compare the particle number at asymptotically early and late times, or with respect to the vacuum states defined in two different frames and do not involve any loop calculation. However, the main problem one encounters when treating quantization in expanding backgrounds concerns the interpretation of the field theory in terms of particles. The absence of Poincar\'e group symmetry in curved space-time leads to the problem of the definition of particles and vacuum states. The problem  may be solved by using the method of the diagonalization of instantaneous Hamiltonian  by a Bogoliubov transformation, which leads to finite results for the number of created particles.

To proceed further, note that Eq. (\ref{nv4}) is a second order differential equation with two independent solutions. Each solution $\chi_{k}$ must be normalized for all times according to
\begin{equation}\label{norm}
W_k(\eta)\equiv \chi_{k}(\eta){\chi^*}'_{k}(\eta)-\chi'_{k}(\eta)\chi^*_{k}(\eta)=-2i\,,
\end{equation}
and also they must satisfy the initial conditions at the time $\eta_i$:
\be
\chi_k(\eta_i)=1/\sqrt{\omega_k(\eta_i)}\,,\hspace{1cm} \chi_k'(\eta_i)=i\sqrt{\omega_k(\eta_i)}\,.\label{inicond}
\ee
These initial conditions select the preferred mode functions which determine the vacuum, or lowest energy state, at a particular moment of time $\eta_i$ \cite{grib,mukh2}.

The Bogoliubov coefficients can be calculated and a straightforward calculation leads to the final expression for the total number of created particles and antiparticles in the $k$ mode \cite{grib,mukh2}:
\begin{equation}\label{Nk}
N_k(\eta)=\bar{N}_k(\eta)={1\over 4\omega_k(\eta)}|\chi'_{k}(\eta)|^2+{\omega_k(\eta)\over 4}|\chi_{k}(\eta)|^2 -{1\over 2}\,.
\end{equation}
The total number density of created particles $n$ is readily obtained by integrating over all the modes \cite{grib,staro}:
\be\label{n}
n(\eta)={1\over 2\pi^2a(\eta)^3 }\int_0^\infty k^2 N_k(\eta) dk\,.
\ee

\section{Particle creation in a $f(R)$ theory}

Now we will apply the above results to the study of scalar particle creation in a radiation dominated universe with a scale factor that follows from a modified $f(R)$ gravity in the Palatini approach, discussed in detail by Allemandi et al \cite{allemandi2}, namely
\be\label{eq14}
f(R)= \alpha R +{\tilde{\beta}\over m-2} R^m\,,\hspace{1cm} m \neq 1,\, 2.
\ee
Such model is exactly that one studied by Carvalho et al. 2008 if we make the correspondence $\alpha =1$, $\tilde{\beta}=-\beta (m-2)$ and $m=-n$. In the metric formalism such model support inflation, while in the Palatini formalism they provide explanation for the present time acceleration \cite{meng}.

The Hubble constant that follows from this $f(R)$ theory is, for radiation dominated universe, (for a detailed treatment of such $f(R)$ model, see Eq. (39) of the Ref. \cite{allemandi2}):
\be
H^2=\Bigg({\dot{a}\over a} \Bigg)^2 = S + L a^{-4}-Ka^{-2}\,,\label{hubble}
\ee
with
\be
S = \epsilon {m-2+\alpha \over 12 \alpha (m-1)}\Bigg({\alpha\over \tilde{\beta}}\Bigg)^{1/(m-1)}\,\,,\hspace{1cm} L = {\kappa \bar{\rho}(m-2)\over 6\alpha (m-1)}\,,\label{SL}
\ee
where $K$ is the spatial curvature ($K=0,\,1,\, -1$), $\kappa = 8\pi G$, $\epsilon = \pm 1$ and  $\bar{\rho}$ is a constant that characterizes the energy density, $\rho = \bar{\rho}a^{-3(1+w)}$, with $w=1/3$ for radiation.

The deceleration parameter can be calculated to be
\be
q={L a^{-4}- S  \over S - K a^{-2} + L a^{-4}}\,.\label{q}
\ee
At early times, in the limit of large energy density, characterized by $\bar{\rho} a^{-4}$, the term with $L$ dominates and we obtain $q \to 1$, a decelerating phase. At late times, on the contrary, the term with $S$ dominates and we obtain $q\to -1$, corresponding to a presently accelerating universe.

At early time, when $La^{-4} >> S$, the solution of (\ref{hubble}) to the scale factor is:
\be
a(t)=\sqrt{2}L^{1/4}t^{1/2}\,.\label{atearly}
\ee
Such solution has exactly the same form of a radiation dominated universe in the standard Einstein cosmology, namely $f(R)=R$, where $a(t)=a_0t^{1/2}$. In terms of the conformal time (\ref{conf}), the expression (\ref{atearly}) turns:
\be
a(\eta)=L^{1/2}\eta \,, \hsp 0 < \eta < \infty \,.\label{aetaearly}
\ee

At late times the term with $S$ in (\ref{hubble}) dominates, and the solution for the scale factor is
\be
a(t)=a_{S}e^{\sqrt{S}\,t}\,,\label{atlate}
\ee
with $a_{S}$ a constant. Such solution has exactly the same form of a de Sitter universe, which is a particular case of a flat homogeneous and isotropic universe with a positive cosmological constant $\Lambda$, namely $a(t)=a_\Lambda e^{H_\Lambda t}$.

In terms of the conformal time, the expression (\ref{atlate}) turns:
\be
a(\eta)=-{1\over \sqrt{S}\,\eta}\,, \hsp -\infty < \eta < 0 \,.\label{aetalate}
\ee

Now let us illustrate the phenomenon of particle creation in this $f(R)$ gravity. We restrict our analysis to a flat universe, $K=0$. 

By analyzing the expression (\ref{m68}), we find $\omega_k^2=k^2 +  M^2 a^2 - a''/a$. Note that the physical wavenumber $k_{ph}$ is related to $k$ by $k_{ph}=k/a$, and its meaning is better understandable in terms of the physical wavelength $\lambda_{ph}$, where $\lambda_{ph}=2\pi / k_{ph}$. Thus we have two limiting cases: $k_{ph}>>M$ (ultrarelativistic particles) and $k_{ph}<<M$ (nonrelativistic particles). 

Here we study the case of nonrelativistic or super-massive particle creation on a radiation dominated universe, so we restrict our results to the early times (\ref{aetaearly}).

For nonrelativistic (or super-massive) particles, $M>>k_{ph}$, at early times we have $\omega_k^2 \approx L M^2\eta^2$, and  the solution of (\ref{nv4}) satisfying the initial conditions (\ref{inicond}) and the normalization (\ref{norm}) is given by:
\be
\chi_{k}(\eta)=A(\eta_i)\sqrt{\eta}\,Y_{1\over 4}(q\eta^2) - B(\eta_i)\sqrt{\eta}\,J_{1\over 4}(q\eta^2)\,,\label{soluc2}
\ee
where $J_\nu$ and $Y_\nu$ are the Bessel functions of first and second kind, respectively, $\nu$ is its order, $q={1\over 2}M L^{1/2}$ and $A$ and $B$ are complex constants depending on the initial time $\eta_i$,
\beqq
A(\eta_i)={J_{1\over 4}(q\eta_i^2)-2q \eta_i^2 J_{5\over 4}(q\eta_i^2)- 2Iq \eta_i^2 J_{1\over 4}(q\eta_i^2)\over M^{3/2}L^{3/4}\eta_i^3[J_{1\over 4}(q\eta_i^2)Y_{5\over 4}(q\eta_i^2)-Y_{1\over 4}(q\eta_i^2)J_{5\over 4}(q\eta_i^2)]  }\,,\nonumber\\
B(\eta_i)= {Y_{1\over 4}(q\eta_i^2)-2q \eta_i^2 Y_{5\over 4}(q\eta_i^2)- 2Iq \eta_i^2 Y_{1\over 4}(q\eta_i^2)\over M^{3/2}L^{3/4}\eta_i^3[J_{1\over 4}(q\eta_i^2)Y_{5\over 4}(q\eta_i^2)-Y_{1\over 4}(q\eta_i^2)J_{5\over 4}(q\eta_i^2)]  }\,.
\eeqq

Note that in the limit of nonrelativistic particles, the solution $\chi$ is independent of the wavenumber $k$, thus the integral in (\ref{n}) to the total number density of created particles diverges to large $k$ (ultraviolet limit). We assume that there is an ultraviolet cutoff $k_{max}$ and perform the integral over the range $0<k<k_{max}$, with $k_{max} \sim M$. The final form of $n(\eta)$ is a complicated function depending on the initial conformal time $\eta_i$, the mass $M$ and on the parameters $\alpha$ and $n$ (see in (\ref{SL}) that it does not depends on $\beta$). By using (\ref{conf}) it is possible to represent the spectrum in terms of the physical time $t$.

In order to compare with the result of the standard Einstein model with a cosmological constant term ($\Lambda CDM$ model), we present in the Fig. 1 the time evolution of the total number density $n$ for the both cases. For the modified gravity model we use the best fit parameters obtained by Carvalho et al \cite{carvalho}, namely $n=0.03$ and $\beta=4.7$ in the range $n \in [-0.25,0.35]$ and $\beta\in [2.3,7.1]$ at 99.7\% c.l., with $\alpha = 1$. We see that the effect of the modified gravity is to increase slighty the total number density of nonrelativistic particles created as compared to the standard $\Lambda CDM$ model. We see that in both cases, after a brief growth on the total number density after the initial time $\eta_i$ the total number density decreases in both cases. We also note that the total number density decreases for lower values of the parameter $n$.

\begin{figure}[tb]
\begin{center}
\epsfig{file=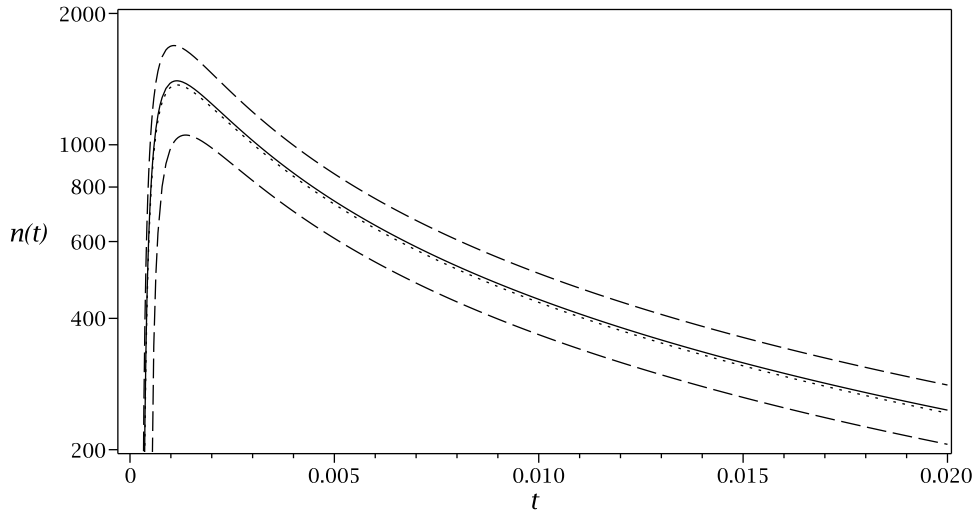, scale=0.67}
\end{center}
\caption{Spectrum of the total number density $n(t)$ of nonrelativistic particles ($M>>k_{ph}$), for $f(R)=R-\beta/R^n$ with $n=0.03$ (solid line), $n=-0.25$ (lower dashed line) and  $n=0.35$ (upper dashed line). The standard $\Lambda CDM$ model is represented in doted line, very close to the best fit value of $n$. The vertical and horizontal scales are arbitrary, corresponding to $\kappa=1$, $\bar{\rho}=100$, $M=10$ and $\eta_i = 0.01$.}\label{fig1}
\end{figure}

\section{Concluding remarks}
In this paper, we have studied the cosmological particle creation process to super-massive (nonrelativistic) real scalar particles in the framework of a $f(R)=R-\beta/R^n$ modified gravity theory. The free parameters $n$ and $\beta$ was constrained by Carvalho et al. 2008 \cite{carvalho}, where the best-fit values for $n$ and $\beta$ obtained from H(z)+BAO+CMB test are $0.03$ and $4.7$, respectively, with the parameters $n$ to lie in the intervals $n \in [-0.25,0.35]$ and $\beta \in [2.3,7.1]$ at 99.7\% c.l.. We have compared the results to the standard $\Lambda CDM$ model, which corresponds to $n=0$ and $\beta = 4.38$, and we have shown that, to super-massive particles, the effects of the modified gravity is to slightly increase the total particle number density as compared to the standard case at early times.  Such mechanism could be an alternative process to create dark matter particles in the universe. This opens the possibility that these particles could give rise to the dark matter in the early universe.


\begin{acknowledgements}
SHP is grateful to CNPq - Conselho Nacional de Desenvolvimento Cient\'ifico e Tecnol\'ogico, Brazilian research agency, for the financial support, process number 477872/2010-7. R.F.L.H thanks INCT-A and is supported by CNPq (No. 478524/2013-7).
\end{acknowledgements}

\end{document}